\documentclass[aps,preprint]{revtex4}
\begin{document}
\draft
\title{\bf Experimental simulation of quantum graphs by microwave
networks
           }
\author{Oleh Hul$^1$, Szymon Bauch$^1$, Prot Pako\'nski$^2$, Nazar
Savytskyy$^1$,\\
Karol {\.Z}yczkowski$^{2,3}$, and Leszek Sirko$^1$}
\address{$^1$Institute of Physics, Polish Academy of Sciences,
Aleja \ Lotnik\'{o}w 32/46, 02-668 Warszawa, Poland \\
$^2$Instytut
Fizyki im. Smoluchowskiego, Uniwersytet Jagiello\'nski, ul.
Reymonta 4, 30-059 Krak\'ow, Poland \\
$^3$Center for Theoretical Physics, Polish Academy of Sciences,
Aleja \ Lotnik\'{o}w 32/46, 02-668 Warszawa, Poland}
%\date{\today}
\date{February 3, 2004}

\bigskip

\begin{abstract}

We present the results of experimental and theoretical study of
irregular, tetrahedral microwave networks consisting of coaxial
cables (annular waveguides) connected by T-joints.
 The spectra of the networks were
measured in the frequency range 0.0001--16 GHz in order to obtain
their statistical properties such as the integrated nearest
neighbor spacing (INNS) distribution and the spectral rigidity
${\Delta_{3}(L)}$. The comparison of our experimental and
theoretical results shows that microwave networks can simulate
quantum graphs with time reversal symmetry (TRS). In particular,
we use the spectra of the microwave networks to study the periodic
orbits of the simulated quantum graphs. We also present
experimental study of directional microwave networks consisting of
coaxial cables and Faraday isolators for which the time reversal
symmetry is broken. In this case our experimental results indicate
that spectral statistics of directional microwave networks deviate
from predictions of Gaussian orthogonal ensembles (GOE) in random
matrix theory approaching, especially for small eigenfrequency
spacing $s$, results for Gaussian unitary ensembles (GUE).
Experimental results are supported by the theoretical analysis of
directional graphs.

\end{abstract}


\bigskip
\maketitle

% {\b {Introduction}}
\smallskip

Quantum graphs of connected one-dimensional wires  were introduced
more than fifty years ago in order to describe organic molecules
by free electron models \cite{Pauling,Kuhn}. They can be
considered as idealizations of physical networks in the limit
where the widths of the wires are much smaller than their lengths,
i.e. assuming that the propagating waves remain in a single
transversal mode. Among the systems modeled by quantum graphs one
can find e.g., electromagnetic optical waveguides \cite{Flesia,
Mitra}, quantum wires \cite{Ivchenko, Sanchez}, mesoscopic systems
\cite{Imry, Kowal} and excitation of fractons in fractal
structures \cite{Avishai, Nakayama}. In spite of the attention
payed to quantum graphs, the statistical properties of their
spectra were hardly investigated in the past. Recently spectral
properties of quantum graphs have been studied in the series of
papers by Kottos and Smilansky
\cite{KottosSmilansky,Kottos,Prlkottos}. They have shown that
quantum graphs are excellent paradigms of quantum chaos. However,
in spite of numerous theoretical investigations of this topic
\cite{Zyczkowski,Kus,Tanner,Tanner2,Kottosphyse,Kottosphysa,Gaspard,Blumel}
no experiments have been performed so far.

The main aim of this work is to demonstrate that using a simple
experimental setup consisting of microwave networks (throughout
the text we also use the names: microwave graphs or circuits) one
may successfully simulate quantum graphs. The circuits are
constructed of coaxial cables (annular waveguides) connected by
T-joints. Furthermore, to mimic the effects of the time reversal
symmetry-breaking in quantum systems it is sufficient to add the
Faraday isolators into the circuit.

The analogy between quantum graphs and microwave networks is based
upon the equivalency of the Schr\"odinger equation
 describing the quantum system and the telegraph equation describing the ideal
 microwave circuit.
 It is worth noting that this paper continues the use of microwave spectroscopy to verify
 wave effects predicted on the basis of quantum physics, which for
 two-dimensional systems, thin microwave cavity resonators, was pioneered by \cite{Stockmann90}
 and further developed by
 \cite{Sridhar,Richter,So,Stoffregen,Haake96,Sirko97}. The first
  microwave experiment specifically devoted to study of quantum chaotic scattering was
  reported in \cite{Doron90}. Later on a similar experimental technique was applied
   in the observation of resonance trapping in an open microwave cavity \cite{Persson00}.
  In the case of two dimensions the
Schr\"odinger
 equation for quantum billiards is equivalent to the Helmholtz equation for microwave cavities
 of corresponding shape. Three-dimensional
 chaotic billiards have been also studied experimentally in the
 microwave frequency domain \cite{Sirko95,Richter2002} but for
 these systems there is no direct analogy between the vectorial Helmholtz
 equation and the Schr\"odinger equation.

     A general microwave graph consists of $N$ vertices (T-joints in our case)
     connected by bonds e.g., coaxial cables.
Following \cite{Kottos} we define the $N\times N$ connectivity
matrix $C_{ij}$ which takes the value $1$ if the vertices $i$ and
$j$ are connected and $0$ otherwise. The coaxial cable consists of
an inner conductor of radius $r_1$ surrounded by a concentric
conductor of inner radius $r_2$. The space between the inner and
the outer conductors is filled with a homogeneous material having
the dielectric constant $\varepsilon$.
 For frequency $\nu$
below the onset of the next TE$_{11}$ mode \cite{Jones}, inside a
coaxial cable can propagate only the fundamental TEM mode, in the
literature often called a Lecher wave. This mode exists because
the cross section of a coaxial line is doubly connected,  what
results in the existence of the potential difference between the
inner and the outer conductors (see Eq. (2)).

     In order to find propagation of a Lecher wave inside the coaxial cable
joining the $i$--th and the $j$--th vertex of the microwave graph
we can begin with the continuity equation for the charge and the
current on the considered cable (bond) \cite{Landau}

\begin{equation}
\label{curcon} \frac{de_{ij}(x,t)}{dt}=-\frac{dJ_{ij}(x,t)}{dx},
\end{equation}
where $e_{ij}(x,t)$ and $J_{ij}(x,t)$ are the charge and the
current per unit length on the surface of the inner conductor of a
coaxial cable.

For the potential difference we can write down
\begin{equation}
\label{eds}
U_{ij}(x,t)=V_{2}^{ij}(x,t)-V_{1}^{ij}(x,t)=\frac{e_{ij}(x,t)}{{\cal
C}},
\end{equation}
where $V_{1}^{ij}(x,t)$ and  $V_{2}^{ij}(x,t)$ are the potentials
of the  inner and the outer conductors of a coaxial cable and
${\cal C}$ is the capacitance per unit length of a cable.

Taking the spatial derivative of (\ref{eds}) and assuming that the
 wave propagating along the cable is monochromatic
$e_{ij}(x,t)=e^{-i\omega t}e_{ij}(x)$ and $U_{ij}(x,t)=e^{-i\omega
t}U_{ij}(x)$ one can obtain \cite{Landau}
\begin{equation}
\label{eds1} \frac{d}{dx}U_{ij}(x)=-{\cal Z}J_{ij}(x),
\end{equation}
where ${\cal Z}={\cal R}-\frac{i\omega {\cal L}}{c^2}$. ${\cal R}$
and ${\cal L}$ denote the resistance and the inductance per unit
length, respectively. The angular frequency $\omega$ is equal to
$2\pi\nu$ and $c$ stands here for the speed of light in a vacuum.

Making use of the equations (\ref{curcon}-\ref{eds1})  and the
definition of ${\cal Z}$ for an ideal lossless coaxial cable with
the resistance
 ${\cal R}=0$, one can derive the telegraph equation on the
 microwave graph
\begin{equation}
\label{wave}
\frac{d^2}{dx^2}U_{ij}(x)+\frac{\omega^2\varepsilon}{c^2}U_{ij}(x)=0,
\end{equation}
where  $\varepsilon={\cal LC}$ \cite{Goubau}.

The continuity equation for the potential difference requires that
for every $i=1,...,N$
\begin{equation}
\label{contin} U_{ij}(x)\vert_{z=0}=\varphi_{i}, \quad
U_{ij}(x)\vert_{z=L_{ij}}=\varphi_{j},
 \quad i<j, \quad C_{ij}\neq 0.
\end{equation}

The current conservation condition
\begin{equation}
\label{currcons}
\sum_{j<i}C_{ij}J_{ji}(x)\vert_{x=L_{ij}}-\sum_{j>i}C_{ij}J_{ij}(x)\vert_{x=0}=0
\end{equation}
may be transformed using Eq. (\ref{eds1}) into
\begin{equation}
\label{currcons1}
-\sum_{j<i}C_{ij}\frac{d}{dx}U_{ji}(x)\vert_{x=L_{ij}}
+\sum_{j>i}C_{ij}\frac{d}{dx}U_{ij}(x)\vert_{x=0}=0,
\end{equation}
where $L_{ij}$ represents the length of the bond joining the
$i$--th and the $j$--th vertex of the graph.  To simplify the
notation, the lengths of six bonds of the four--vertex graph shown
in Figure~1 are labeled by letters $\{a,...,f\}$.

 Assuming the following correspondence:
$\Psi_{ij}(x)\Leftrightarrow U_{ij}(x)$ and $k^2\Leftrightarrow
\frac{\omega^2 \varepsilon}{c^2}$,
 equation (\ref{wave}) is formally equivalent
     to the one--dimensional Schr\"{o}dinger equation (with $\hbar=2m=1$) on the graph
 with the magnetic vector potential $A_{ij}=0$ \cite{Kottos},

\begin{equation}
\label{Schred} \frac{d^2}{dx^2}\Psi_{ij}(x)+k^2\Psi_{ij}(x)=0.
\end{equation}

It is easy to verify that equations (\ref{contin}) and
(\ref{currcons1}) are equivalent  to equations derived in
\cite{Kottos}  (see Eqs. (3)) for quantum graphs with Neumann
boundary conditions ($\lambda_{i}=0$) and vanishing magnetic
vector potential $A_{ij}=A_{ji}=0$. Such systems  possess the time
reversal symmetry (TRS),

\begin{eqnarray}
\label{bound} \left\{ \begin{array}{cc}
\Psi_{ij}(x)\vert_{x=0}=\varphi_{i}, \quad
\Psi_{ij}(x)\vert_{x=L_{ij}}=\varphi_{j}, \quad i<j, \quad
C_{ij}\neq 0
\\
\\
-\sum_{j<i}C_{ij}\frac{d}{dx}\Psi_{ji}(x)\vert_{x=L_{ij}}+
\sum_{j>i}C_{ij}\frac{d}{dx}\Psi_{ij}(x)\vert_{x=0}=0.
\end{array}\right.
\end{eqnarray}

%{\b {Experiment}}
\smallskip
In order to check the equivalence of microwave and quantum graphs
we measured the spectra of ten tetrahedral microwave graphs in the
frequency range $0.0001-16$ GHz. For this frequency range only a
Lecher wave can propagate in the graphs. The next mode to
propagate is the TE$_{11}$ with the cut-off frequency $\nu_{c}
\simeq \frac{c}{\pi (r_1+r_2)
 \sqrt{\varepsilon}} = 32.9$ GHz
\cite{Jones}, where $r_1$ = 0.05 cm is the inner wire radius of
the coaxial cable (SMA-RG402), while $r_2$ = 0.15 cm is the inner
radius of the surrounding conductor, and $\varepsilon$ = 2.08 is
Teflon dielectric constant.

The experimental set-up for measurements of spectra of the
microwave graphs is shown in Figure~1. We used Hewlett-Packard
8672A microwave synthesizer to measure the spectra of the graphs
in the frequency range 2 - 16 GHz while for the frequency range
0.0001 - 3 GHz Rhode-Schwartz SMT-03 microwave synthesizer was
used. The microwave coupler (Narda 4055) enabled us to observe
signals reflected from microwave graphs.  Because  for the
specified frequency range only a single TEM mode could propagate
in the microwave networks a reflected signal was proportional to
$|S|^2$, where the complex number $S$ may be considered as a
one-dimensional scattering matrix. This type of microwave
experiments, related to scattering matrix measurements, was
pioneered by \cite{Doron90} and stimulated by \cite{Blumel88}. In
Figure~2(a) a typical fragment of a measured modulus of scattering
matrix $|S|$ of the graph is presented in the frequency range 3.95
- 5.05 GHz. The experimental spectrum is also compared with
numerically calculated eigenfrequencies of the ideal graph ($R=0$)
having the same bonds lengths as the experimental one. The total
``optical" lengths of the microwave graphs, including T-joints,
varied from 171.7 to 262.2 cm, which allowed for the observation
of 156-264 eigenfrequencies in the frequency range 0.0001 - 16
GHz.
 To avoid the degeneracy of eigenvalues
the lengths $L_{i,j}$ of the bonds (cables) were chosen not to be
commensurable.  The transmission through the T-joints was
characterized by the weak frequency dependence e.g. in the
frequency range 0.05 -16 GHz the ratio
$R_S=(|S_{ij}|^{max}-|S_{ij}|^{min})/|S_{ij}|^{max} \leq 0.06$,
where $|S_{ij}|^{max}$ and $|S_{ij}|^{min}$ are the maximum and
the minimum values of modulus of non-diagonal elements of a three
port scattering matrix $S_{ij}$, respectively. The indices
$i,j=1,2,3$ and $i \neq j$. For the frequency range 3.95-5.05 GHz
specified in Figure~2(a) the ratio fulfilled the condition $ R_S
\leq 0.02$.

  For more comprehensive comparison of the experimental and
numerical results the experimental spectrum shown in Figure~2(a)
is compared in Figure~2(b) with the response function $r(k)$
calculated for the graph having the same bonds lengths as the
experimental one. The response function $r(k)$ was introduced in
this paper in order to analize the directional graphs consisting
of Faraday isolators and is defined by Eq. (\ref{RK}). In the
calculations of the response function  $r(k)$  absorption of
microwave cables were taken into account by replacing the real
wave vector $k$ by the complex vector $k+i\beta\sqrt{k}$
\cite{Goubau}. The absorption coefficient $\beta=0.009$ $m^{-1/2}$
was evaluated on the basis of absorption of the microwave cables
used in the experiment. The direct comparison of the results
presented in Figure~2(a) and Figure~2(b) requires some care
because the response function $r(k)$ is rather proportional to the
amplitude of the field transmitted through the graph
 than to the amplitude reflected from the graph that is represented by the
 scattering matrix $|S|$.
However, the aim of this comparison is to show that the inclusion
of absorption of microwave cables leads to comparable with the
experimental results broadening of resonanses. It is also
important noting that calculated in such a way eigenfrequencies
are very close to the ones calculated for the ideal graph, from
which they differ at most by 1 MHz.
 Figures~2(a-b) show that the
agreement between experimental and theoretical results is quite
good (the relative errors are of the order of $10^{-3}$), what
justifies a posteriori our assumption that the microwave circuits
can be described with good accuracy by the quantum graphs with
Neumann boundary conditions.  In this way our experimental results
additionally support theoretical findings about the boundary
conditions for shrinking domains \cite{Kuchment,Rubinstein}. Our
results also show that relatively short microwave graphs
consisting of coaxial cables, at least as it concerns the
eigenvalues positions, can be approximately treated as ideal
lossless graphs with the resistance ${\cal R}=0$. The last
statement is not very surprising.  A similar situation one can
find in the experiments with microwave cavities
\cite{Stockmann90}.  A thorough discussion on influence of
absorption of energy caused by the finite conductivity of the
cavity walls on reflected from the cavity power was given by Doron
et al \cite{Doron90}. They showed (see also \cite{Persson00}) that
small absorption of energy is necessary for revealing cavity's
resonances as dips in the reflected power.

     We have examined statistical properties of spectra of the microwave graphs
      such as the integrated nearest neighbor
spacing (INNS) distribution $I(s)$   and the spectral rigidity
${\Delta_{3}(L)}$. (for their definitions see eg. \cite{Haake,
Stockmann}). Figure~3  presents the INNS distributions. The solid
line represents predictions of random matrix theory obtained for
Gaussian Orthogonal Ensemble (GOE), applicable for systems with a
time--reversal symmetry. The dashed line denotes results
characteristic of Gaussian Unitary Ensemble (GUE), used if the
time reversal symmetry is broken \cite{Haake}. Experimental curve
(open triangles) was obtained by averaging over the set of $10$
microwave graphs obtained by varying  with the length of one bond,
which provided us with the total of 2220 experimentally measured
eigenfrequencies. Numerical curve (open circles) shows results
averaged for ten quantum graphs having the same bonds lengths as
the experimental ones. Eigenfrequencies were calculated by solving
numerically the secular equations for quantum graphs (Eqs. (6-8)
in \cite{Kottos}). This procedure allowed us to identify total of
2344 eigenfrequencies, slightly more then measured experimentally.
Figure~3 shows that in both cases the INNS distributions are in a
very good agreement with the GOE predictions.

Figure~4 demonstrates the spectral rigidity ${\Delta_{3}(L)}$
obtained for the microwave graph of the ``optical" length 223.6
cm. Experimental curve (open triangles) was based on 229
identified eigenfrequencies, while the numerical data (open
circles) were computed out of 237 eigenfrequencies.
 In both cases the frequency range was 0.0001--16 GHz.
 Deviations of the experimental and
numerical rigidity from the GOE predictions (solid line) are
visible. For comparison the dashed line in Figure~4 shows the RMT
prediction for GUE. Our experimental and numerical results are
lying above the GOE prediction for  $L$ between $2.5-5$. For
higher value of $L$ a saturation of the numerical value of the
spectral rigidity is observed in agreement with the predictions of
Berry \cite{Berry}. The experimental rigidity for $L>10$ is
located below the GOE curve and above the numerical results. The
departure of the experimental rigidity from the numerical one can
be probably attributed to the loss of about 3\% of experimental
eigenfrequencies.

The measurements of the spectra of the graphs enabled us also to
calculate the lengths of periodic orbits in the graph.
 They were computed from the Fourier transform

\begin{equation}
F(l)=\int_{0}^{k_{max}}\tilde{\rho}(k)\omega(k)e^{-ikl}dk,
\end{equation}
where $\tilde\rho$ is the oscillating part of the level density
and $\omega(k)=sin^2(\pi\frac{k}{k_{max}})$ is a window function
that suppresses the Gibbs overshoot
phenomenon~\cite{Sirko97,Bauch}. Here $k_{max}$ is the  maximal
value of the wave number within the interval where  the
eigenvalues of the graph were evaluated. In order to extract the
oscillating part of the level density $\tilde{\rho}$ we determined
the density of states according to $\rho(k) = \sum_j
\delta(k-k_j)$ and subtracted from it  the mean density $\bar
\rho(k) = d\bar N(k)/dk $.  The mean $\bar N(k)$ of the staircase
function, i.e.,  the number of resonances up to the wave number
$k$, was obtained from a least squares fit $\bar N(k) = \alpha_1 k
+ \alpha_2$ of the measured staircase $N(k)$. The slope parameter,
obtained from the experimental data, $\alpha_1=0.707 \pm 0.006$,
is very close to the value $\alpha=0.712$, received from the Weyl
formula given by Eq. (7) in \cite{KottosSmilansky}.

The absolute square of the Fourier transform of the fluctuating
part of the density of resonances $|F(l)|^2$  for the graph of the
``optical" length 223.6 cm is shown in Figure~5. The lengths of
the bonds of this graph fulfill the following relations:
$a<b<c<d<e<f$. Results obtained from the experimental spectrum
(solid line) are compared to the results obtained from numerical
calculations (dotted line). The experimental spectrum included 149
identified eigenfrequencies while the numerical one  150
eigenfrequencies. In both cases the frequency range was 0.2--10.2
GHz. We used the narrower frequency range than in the calculations
of the INNS distributions and ${\Delta_{3}(L)}$ to be sure that at
most only one eigenfrequency was missing in the experimental
spectrum. The absolute square of the Fourier transform $|F(l)|^2$
shows pronounced peaks near the lengths of certain periodic
orbits. The agreement between the experimental and the numerical
results is good, however, some discrepancies for shorter periodic
orbits are visible. In Figure~5 we show all irreducible periodic
orbits \cite{Kottos}, i.e. periodic orbits which do not intersect
themselves, with the lengths $l<165$ cm. For clarity, in Figure~5
we additionally show the first repetition of the periodic orbit
$2b$ at $l=2b+2b=105.2$ cm and two reducible periodic orbits
$abc+2c$ and $bde+2a$ at $l=149.3$ cm and $l=154.7$ cm,
respectively. It should be noticed that many peaks for $l>70$ cm
cover several unresolved periodic orbits.  In order to check
whether the missed resonance was responsible for discrepancies in
the lengths of periodic orbits we artificially added this
resonance to the spectrum of the graph and recalculated the
lengths of periodic orbits (results are not shown). Indeed, the
inclusion of the missed level improved the agreement with the
numerical results. The main change was visible in the
"experimental" amplitudes that became closer to the numerical
ones. For example the amplitudes of the most sensitive orbits $2a$
and $2b$ (see Fig. 5) were decreased from 1.5 to 1.0 and from 1.4
to 1.1, respectively.  The positions of the periodic orbits were
not as sensitive as amplitudes. For example in the case of the
orbit $2a$ the length was changed from $\ell =42.0$ cm to
$\ell=41.4$ cm becoming more distant from the numerical result
$\ell=42.0$ cm. In a different way behaved the orbit $2b$ which
length was modified from $\ell =51.5$ cm to $\ell=52.5$ cm
becoming closer to the numerical result $\ell=52.6$ cm.

In this paper we also present experimental study of  microwave
graphs consisting of coaxial cables and Faraday isolators. The
graphs with Faraday isolators are examples of simple experimental
realization of directional graphs for which the time reversal
symmetry is broken. A microwave Faraday isolator is a passive
device, which transmits the wave moving in one direction while
absorbing the wave moving in the opposite direction.  Due to
absorption the introduction of Faraday isolators transforms the
problem from the bound system to an open system.  In the
experiment AerCom 60583 Faraday isolators (insertion loss $< 0.4$
dB, isolation $> 19$ dB, length $=5.7$ cm ) with the operating
frequency range $3.5- 7.5$ GHz were used. We measured the spectra
of four graphs consisting in one of their bonds one Faraday
isolator or two Faraday isolators connected in series. The
limitations imposed by the narrow range of isolators operating
frequency lead to rather poor eigenfrequencies statistics -
between 34 and 39 eigenfrequencies were observed  for the graphs
with Faraday isolators. Therefore, for each of four graphs we
performed three measurements where: the isolator was mounted in
the bond $b$ of the graph,  the isolator was mounted in the bond
$d$, and two isolators  connected in series were mounted in the
bond $d$. The assignment of the letters to the bonds of the graph
is shown in Figure~1.

 The
results of these twelve measurements (together 444
eigenfrequencies) were averaged to obtain the INNS distribution
(solid triangles in Figure~6). The INNS distribution obtained for
the same frequency range, but without Faraday isolators are also
shown in Figure~6 (open triangles). The examination of the INNS
distribution obtained for the graphs, with and without Faraday
isolators, shows that they are different. In spite of some
deviations, one can see that INNS distribution for the graphs
without the isolators is close to the RMT prediction for GOE
(solid line) in contrast to the INNS distribution for the graphs
with the isolators, which follows more closely the RMT prediction
for GUE (dashed line). This is especially well seen at small
eigenfrequency spacing $s$. Similar deviations of the spectral
statistics were reported by experiments with microwave billiards
\cite{So,Stoffregen,Haake96}. In the experiment performed by So et
al. \cite{So} the transition from GOE to GUE statistics was caused
by a piece of magnetized ferrite placed inside a two-dimensional
microwave cavity while in the experiment performed in Marburg
\cite{Stoffregen,Haake96} the deviation from GOE statistics was
induced by Faraday isolator connected to a microwave cavity.

The directed graphs can be also modeled theoretically. The crucial
element of the graph --- the Faraday isolator in a directed bond
--- can be described by means of a filter factor which damps the
wave moving in one direction. In the numerical analysis of the
directed graphs we have assumed that measuring the reflection
spectra of the graphs we are rather probing the graphs as  closed
systems by some coupling, which is weak enough, not to influence
the internal dynamics of the systems.

We introduce the connectivity matrix $D$ of a directed graph, which does
not need to be symmetric. With any directed graph $\Gamma$ one may
associate a bidirectional graph $G$ with the same number of vertices.
Its connectivity matrix $C$ is symmetric,
\begin{equation}
\label{CG}
C_{ij} = \max(D_{ij},D_{ji}).
\end{equation}
The number $B$ of bonds in the graph $G$ is equal to $B = \frac{1}{2}
\sum_{ij} C_{ij}$. To be able to filter some waves propagating in one
direction while preserving those moving in the opposite direction we
will use a bond scattering technique of analyzing spectra of graphs,
similar to this introduced by Kottos and Smilansky~\cite{Kottos}.
Consider a plain wave $\Psi_{j' n}(x)=e^{-ikx}$ coming from the vertex
$j'$ to the vertex $n$. It is scattered into all bonds going out from
the vertex $n$, for which $C_{nj} \neq 0$,
\begin{equation}
\label{wavefunction}
 \Phi_{nj}(x)=\delta_{jj'}e^{-ikx}+\sigma_{jj'}^{(n)}e^{ikx}.
\end{equation}
The vertex scattering matrix $\sigma_{jj'}^{(n)}$ is completely
determined, if we assume Neumann boundary conditions~(\ref{bound}),
which imply
\begin{equation}
\label{sigma}
 \sigma_{jj'}^{(n)}=C_{j'n}C_{nj}(-\delta_{jj'}+2/v_n).
\end{equation}
Here $v_n$ denotes the number of bonds meeting at the $n$-th vertex,
also called the \emph{valency} of the vertex. Elements of
$\sigma_{jj'}^{(n)}$ for all vertex $n$ combine to the entire bond
transition matrix of the graph $G$
\begin{equation}
\label{transition}
 T_{jl,nm}=\delta_{ln}C_{jl}C_{nm}\sigma_{jm}^{(l)} \ ,
\end{equation}
which describe the changes of amplitudes of waves propagating in
each bond of the graph (in both directions) after one event of
scattering on vertices. The matrix dimension is equal to twice the
number of bonds $B$ in the graph. To take into account the
presence of the Faraday isolators we make use of the connectivity
matrix $D$ of the directed graph $\Gamma$, and introduce a
diagonal $2B \times 2B$ matrix $\Lambda(k)$
\begin{equation}
\label{lambda}
  \Lambda_{jl,j' l'}(k) =
    \delta_{jj'}\delta_{ll'}D_{jl}e^{ikL_{jl}} \ ,
\end{equation}
where the phase factor describes the free propagation along the
bond $(jl)$ of length $L_{jl}$. By definition, the element
$D_{jl}$ is equal to zero for bonds which do not belong to the
directed graph $\Gamma$.  The effect of absorption of the
microwave field by microwave cables can be easily taken into
account by modifying the matrix $\Lambda_{jl,j' l'}(k)$ given by
Eq. (\ref{lambda}) to the form
$\delta_{jj'}\delta_{ll'}D_{jl}e^{(ik-\beta \sqrt{k})L_{jl}}$,
where $\beta$ is the absorption coefficient.  The total evolution
of the vector of wave amplitudes of length $2B$ is given by the
bond scattering matrix
\begin{equation}
\label{scattering}
  S(k) = \Lambda(k) \cdot T.
\end{equation}
The matrix $S(k)$ is sub-unitary, since it is obtained by putting
to zero some elements of a unitary matrix. We denote the
eigenvalues of $S(k)$ by $\lambda_j(k)$, all of $\lambda_j(k)$ are
located in (or at) the unit circle, $|\lambda_j(k)| \le 1$. The
equation for the eigenmodes of the quantum graph
\begin{equation}
\label{det}
  \det(S(k)-1)=0
\end{equation}
may have no real solution. In our experimental setup the graph is
driven by the microwave generator. We are interested, for which
wave vectors $k$ the resonant driving of the graph will appear. We
analyze the stationary state of the system, in which an arbitrary
number $p$ of scattering processes take place and decompose it in
the eigenbasis of $S(k)$. The amplitudes of each mode become an
infinite superposition of waves scattered $p$ times, so the
enhancement factor $r_j(k)$ of the $j$-th mode reads
\begin{equation}
  r_j(k) = \sum_{p=0}^\infty [\lambda_j(k)]^p = \frac{1}{1-\lambda_j(k)}
  \ ,
\end{equation}
where $\lambda_j(k)$ are the eigenvalues of the bond scattering
matrix $S(k)$.  Since each eigenmode may contribute to the
resonant dissipation in the system, we approximate the total
response function of the graph by the average enhancement factor
(the mean of $r_j(k)$)
\begin{equation}
  \label{RK}
  r(k) = \frac{1}{2B} \; \sum_{j=1}^{2B} r_j(k) = \frac{1}{2B} \;
    \sum_{j=1}^{2B} \frac{1}{1-\lambda_j(k)} \ .
\end{equation}
Maxima of this function, which occur if one of the eigenvalue
$\lambda_j(k)$ is close to unity, identify resonant values of the wave
vector $k$. We analyzed the function $r(k)$ generated for parameters
of the system as used in the experiment and studied numerically the
statistics of its maxima.

Using this approach we calculated approximated eigenfrequencies of
twenty directed graphs in the frequency range 0--20 GHz. As in the
experimental realization, only one bond was assumed to be
directed. Numerical search for the eigenfrequencies  was performed
for two sets of directed graphs with five different lengths of a
directed bond $b$ (see Figure~1 for the assignment of the letters
to the bonds of the graph). The other bonds of the graphs were
bidirectional and within the set were kept fixed. The same number
of numerical calculations were also done for two sets of directed
graphs with the varied length of the directed bond $d$. Figure~7
shows the integrated nearest neighbor spacing distribution
averaged for twenty realizations of the directed graphs in the
frequency range 0 -- 20 GHz. Together 3207 eigenfrequencies were
used in the calculations of the INNS distribution. In this case we
decided not to put the experimental and theoretical data on one
plot, to emphasize that the results are based on different
statistics and cannot be directly compared. However, it is
justified to compare these numerical results obtained for the
directed graphs (solid circles) with the numerical data obtained
in the frequency range 0 -- 20 GHz for twenty realizations of
standard (bidirectional) graphs (empty circles). In this case 4641
eigenfrequencies were used in the calculations of the INNS
distribution. Theoretical predictions for GOE and GUE, denoted by
solid and dashed curve, respectively, suggest that the INNS
spectral statistics for directed graphs deviate at small spacings
from the GOE curve and become closer to the GUE predictions. This
result confirms also our experimental findings for the microwave
directed graphs.

%{\b {Conclusions}}

In summary, we show that quantum graphs with Neumann boundary
conditions can be simulated experimentally by  microwave networks.
Bidirectional microwave graphs, i. e. circuits without Faraday
isolators, simulate quantum graphs with time reversal symmetry.
The results for the directional microwave graphs with Faraday
isolators, for which the time reversal symmetry is broken,
 indicate that
their certain characteristics such as the integrated nearest
neighbor spacing distribution can significantly differ from the
RMT prediction for GOE, approaching the results characteristic of
GUE.

Acknowledgments.  This work was partially supported by KBN grants
No 2 P03B 023 17 and 2 P03B 047 27. We would like to thank
Professors Marek Ku\'s and Petr {\v S}eba for valuable
discussions.

\pagebreak

%\centerline {\bf Figure Captions}

\smallskip

\begin{figure}[!]
\caption{Experimental set-up for measurements of the spectra of
the microwave graphs. Microwave synthesizer: HP8672A (2-18.5 GHz)
and SMT03 (5 kHz-3 GHz), D - crystal detector (HP8472B), C -
microwave coupler (Narda 4055).}\label{Fig1}
\end{figure}

\begin{figure}[!]
\caption{(a) A fragment of a measured modulus of scattering matrix
$|S|$  of the microwave graph of the ``optical" length 223.6 cm in
the frequency range 3.95--5.05 GHz. (b) The response function
$r(k)$ calculated for the graph having the same bonds lengths as
the experimental one.  In the calculations of the response
function $r(k)$ absorption of the microwave field by coaxial
cables were taken into account (see text). Vertical broken lines
show the positions of numerically calculated eigenfrequencies of
the graph without absorption.}\label{Fig2}
\end{figure}

\begin{figure}[!]
\caption{Integrated nearest neighbor spacing (INNS) distribution
$I(s)$
 averaged for 10 microwave graphs. Results of the experiment
(open triangles) are compared with the numerical results (open
circles) and theoretical prediction for GOE (solid line) and GUE
(dashed line).}\label{Fig3}
\end{figure}

\begin{figure}[!]
\caption{Spectral rigidity $\Delta_3(L)$ for the microwave graph
of the ``optical" length $223.6$ cm. Results of the experiment
(open triangles) are compared with the numerical results (open
circles) and theoretical prediction for GOE (solid line) and GUE
(dashed line).}\label{Fig4}
\end{figure}

\begin{figure}[!]
\caption{Absolute square of the Fourier transform    of the
fluctuating part of the density of resonances of the graph of the
``optical" length $223.6$ cm. Results of the experiment (solid
line) are compared with the numerical results (dotted line). The
assignment of peaks of $|F(l)|^2$ to simple periodic orbits (see
text) is shown along with the length of the orbits. The ``optical"
lengths of the bonds of the graph: $a=21.0$ cm, $b=26.3$ cm,
$c=34.0$ cm, $d=39.6$ cm, $e=46.8$ cm, $f=55.9$ cm. }\label{Fig5}
\end{figure}

\begin{figure}[!]
\caption{Integrated nearest neighbor spacing distribution
  averaged for eight realizations of the microwave graphs
 with Faraday isolators (solid triangles) is compared with the
 averaged results for the microwave graphs without the isolators (open
triangles) and theoretical prediction for GOE (solid line) and GUE
(dashed line). In both cases experimental results were obtained
for the frequency range 3.5 -- 7.5 GHz. }\label{Fig6}
\end{figure}

\begin{figure}[!]
\caption{Numerically calculated integrated nearest neighbor
spacing (INNS) distributions averaged for twenty realizations of
the directed graphs (solid circles) is compared with the averaged
results for the bidirectional graphs (open circles). Calculations
were performed in the frequency range 0 -- 20 GHz.
  Numerical results for the INNS distributions are
compared with theoretical predictions for GOE (solid line) and GUE
(dashed line).}\label{Fig7}
\end{figure}

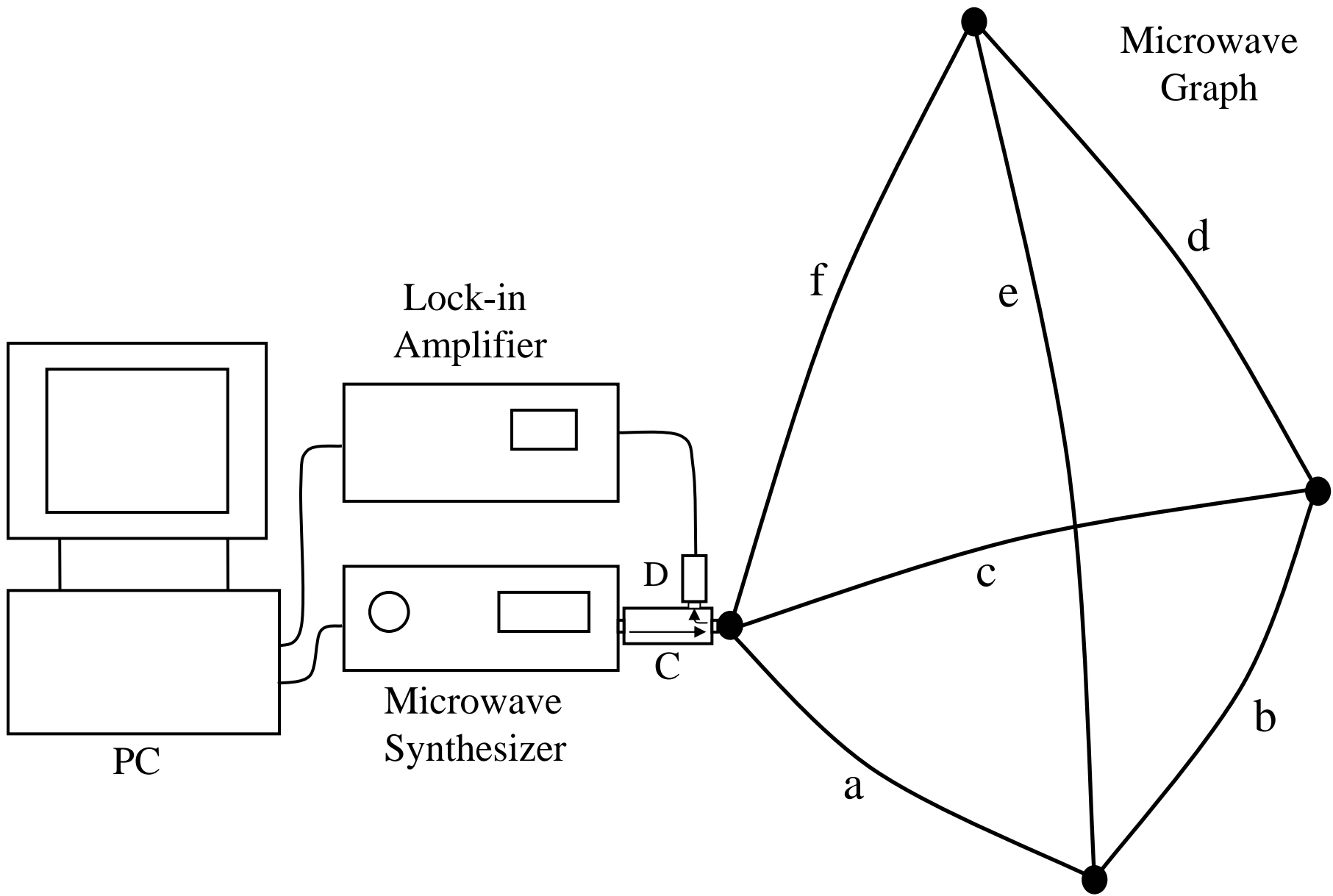
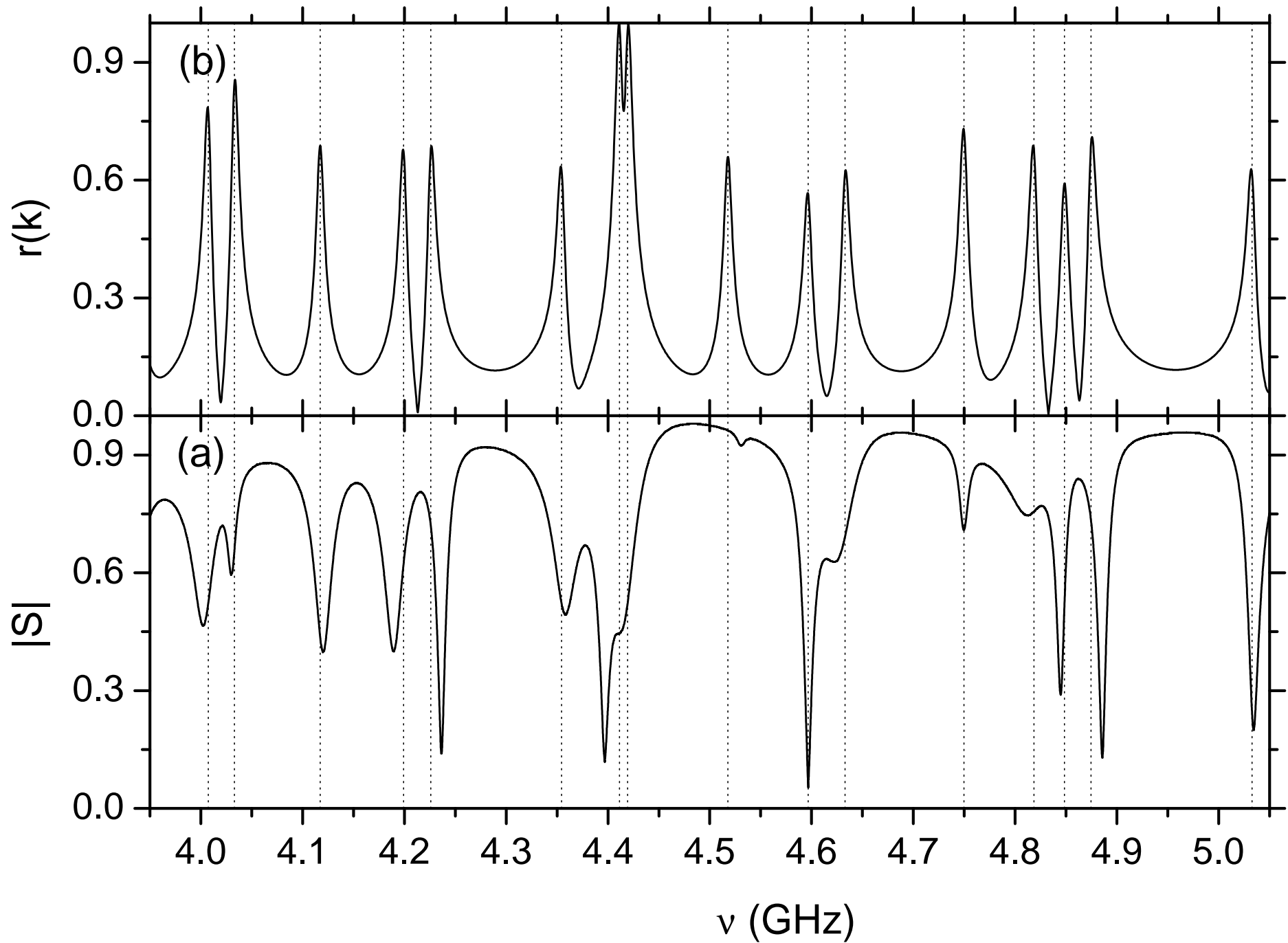
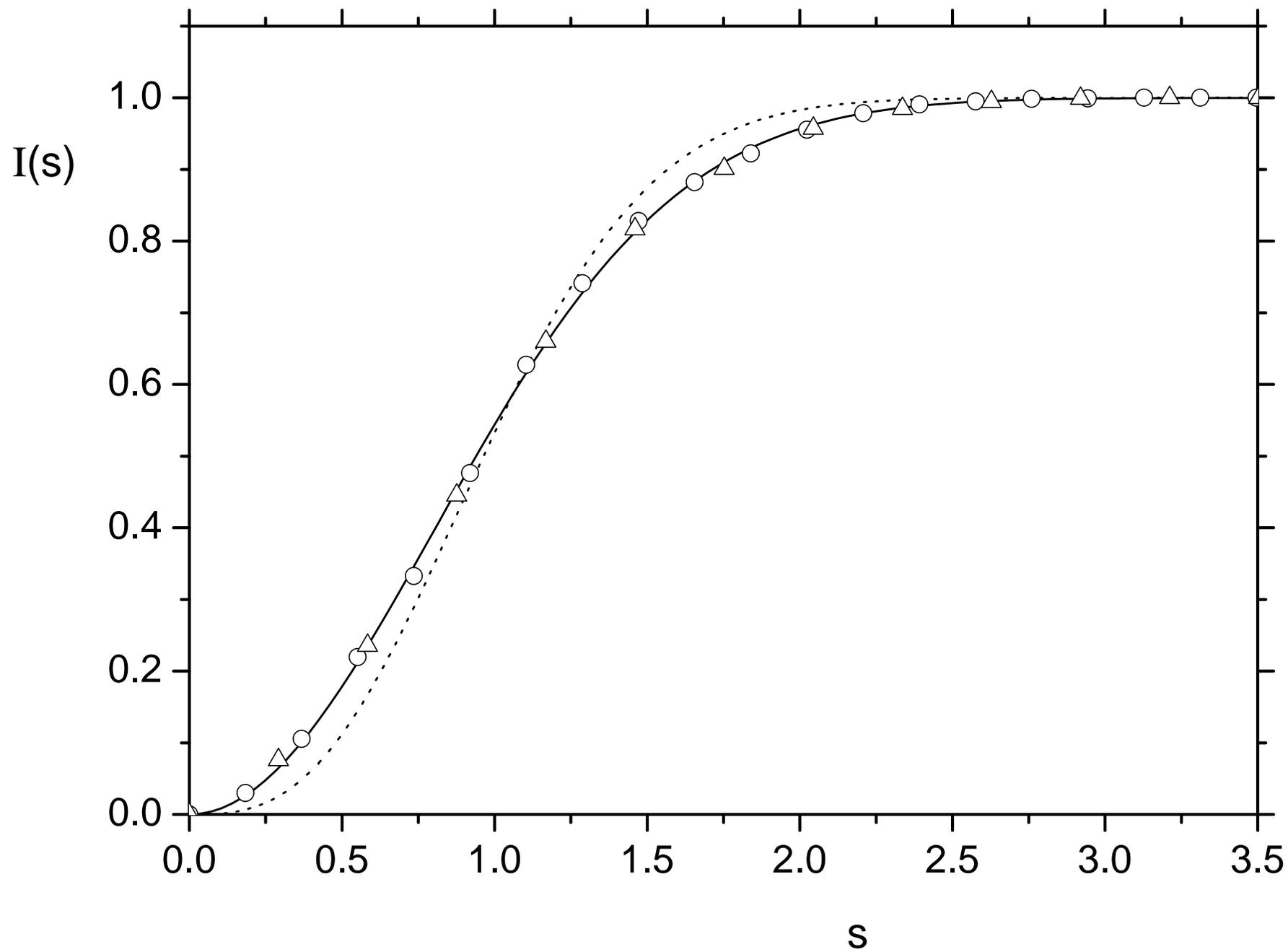
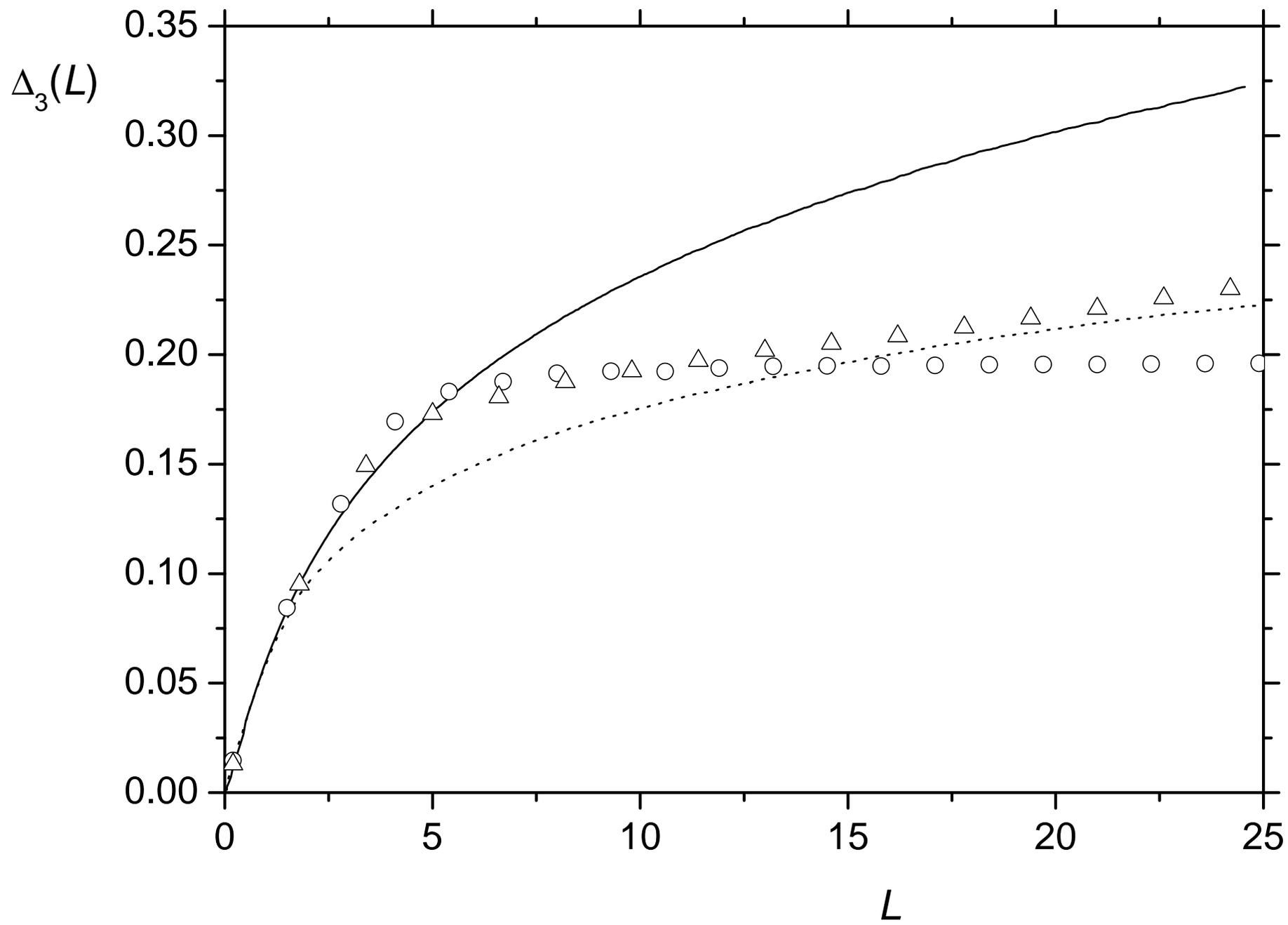
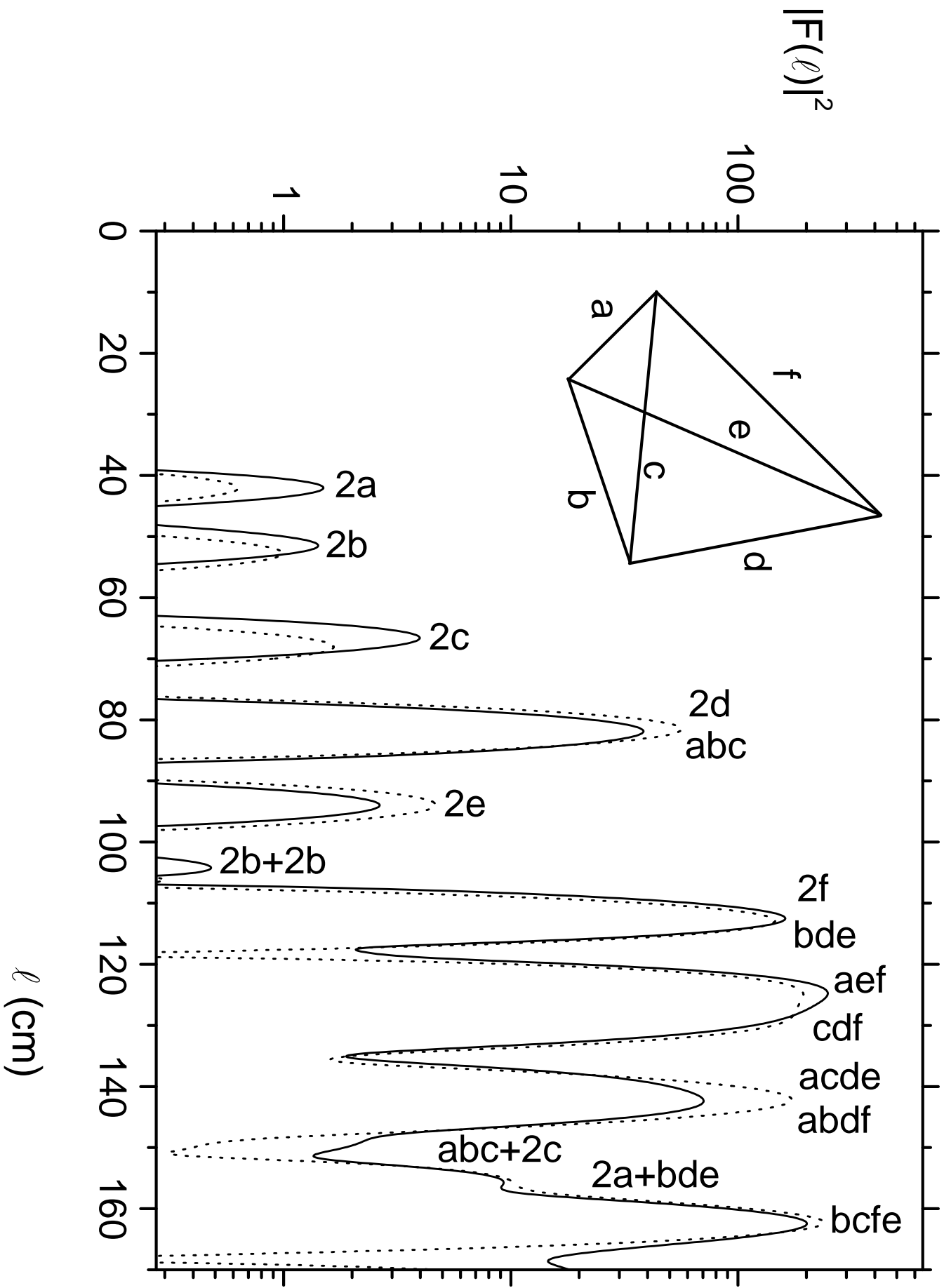
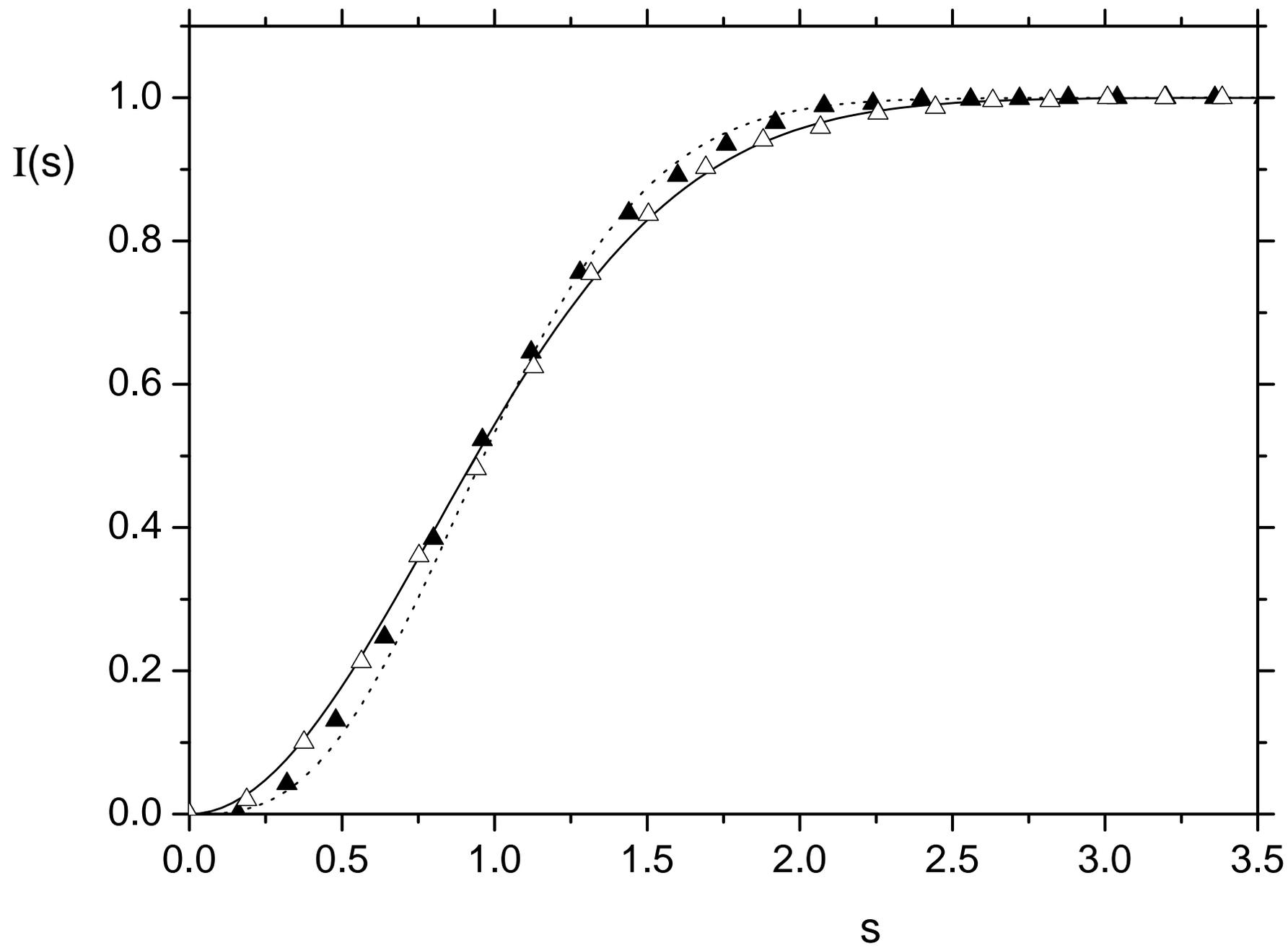
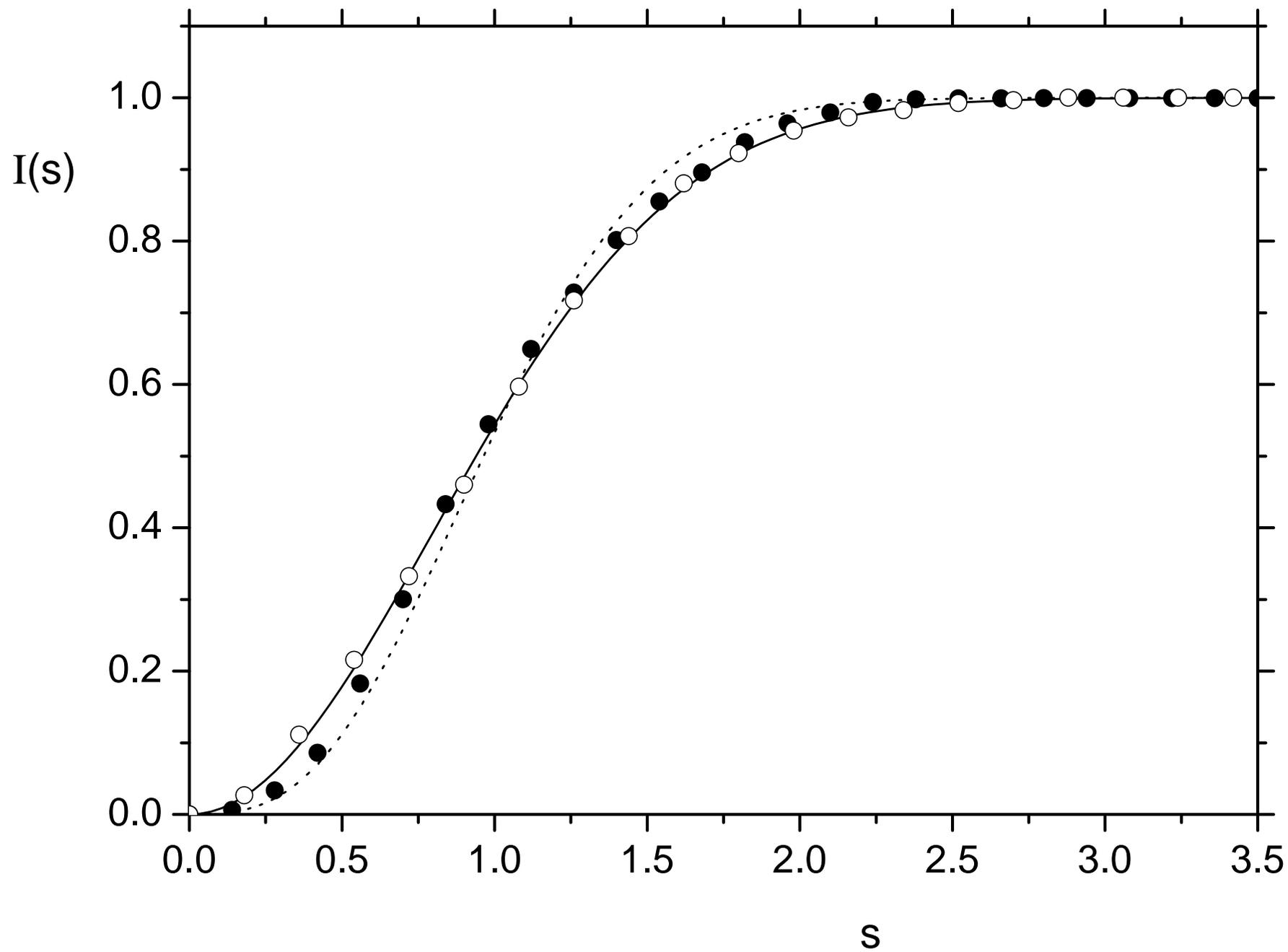
\end{document}